\documentstyle[preprint,aps,epsf,floats]{revtex}

\begin{document}

\preprint{\tighten \vbox{\hbox{CALT-68-2271} 
                \hbox{CMU-HEP00-02}
                \hbox{NSF-ITP-00-38} }}

\title{Extracting $|V_{ub}|$ from the Hadronic Mass Spectrum of \\
Inclusive $B$ decays}

\author{Adam K.\ Leibovich$^a$, Ian Low$^{a,b}$, 
         and I.\ Z.\ Rothstein$^{a,c}$}

\address{\vspace{.5cm}
 $^a$ Department of Physics, 
Carnegie Mellon University, Pittsburgh, PA 15213 \\
$^b$ Institute for Theoretical Physics, 
University of California, Santa Barbara, CA 93106\\
$^c$ California Institute of Technology, Pasadena, CA 91125}

\maketitle

{\tighten
\begin{abstract}
Following a strategy introduced earlier by the authors, we show that
it is possible to extract $|V_{ub}|$ from the cut hadronic mass
spectrum of $B$ decays without large systematic errors which usually
arise from having to model the Fermi motion of the heavy quark.  We
present a closed form expression for $|V_{ub}|/|V_{ts}|$ which is
accurate up to corrections of order $\alpha_s^2$, $\alpha_s \rho$,
$(\Lambda/m_b)^2/\rho$, where $\rho$ is the experimental cut
$(s_c/m_b^2)$ on the hadronic mass used to veto charmed decays. Modulo
duality violation errors, which are intrinsic to all inclusive
predictions, we estimate the theoretical error in the extraction to be
at the $5\%$ level.
\end{abstract}
}


\newpage

\section{Introduction}

An accurate determination of $V_{ub}$ would fill a gaping hole in our
understanding of the Cabibbo-Kobayashi-Maskawa (CKM) sector of the
standard model. Indeed, while it is not clear that we will be able to
cleanly determine the angles of the unitarity triangle, an accurate
measurement of $|V_{ub}|$ could determine whether or not the CKM
description is correct by measuring the lengths of the sides.

While in principle one could measure $|V_{ub}|$ quite easily in a
systematic approximation to QCD\cite{CGG} from an inclusive measurement 
of the uncharmed semi-leptonic branching ratio, in practice, it
seems to be much more difficult.  The problem lies in rejecting the
charmed decay background. This cut causes the canonical expansions in
$\alpha_s$ and $\Lambda/m_b$ to breakdown. The imposition of the cut
changes the expansion parameters to $\alpha_s \log \rho$ and
$\Lambda/(m_b\, \rho)$, where $\rho$ parameterizes the cut and is
numerically small. It may be the case that one or both of these
parameters can be of order 1, thus necessitating a reorganization of
the expansion.  Here we will not review the systematics of these
expansions, as they have been thoroughly discussed in the literature
\cite{AR,LR,LLRI}.

The nature of the abovementioned series depends on the kinematic
variable with which we choose to cut. Each choice has advantages and
disadvantages.  The cut lepton energy spectrum,
$E_l>(M_B^2-M_D^2)/(2M_B)$, is relatively simple experimentally since
there is no need for neutrino reconstruction. Theoretically, this
choice of cuts is challenging due to the breakdown of the
non-perturbative series in $\Lambda/(m_b-2E_l)$. Furthermore, the
fraction of the $b\rightarrow u$ transitions included after the cut is
only about $10\%$.  The cut hadronic invariant mass spectrum is more
challenging from an experimental viewpoint, but has its virtues\cite{D}. First
of all, while this observable smears over the same range of hadronic
masses as the cut lepton spectrum, it is weighted more towards states
of larger invariant mass. We thus would expect local duality to be
more effective in this case. Furthermore, this cut rate includes
around $40-80\%$ of the $b\rightarrow u$ transitions \cite{FN}. In
both of these cases the non-perturbative series must be resummed. The
necessity for resummation of the perturbative series can only be
determined a posteriori. For example, in the case of radiative decay
with an experimental cut of $2.1$ GeV on the photon energy, it was
found \cite{LR} that the resummation of the perturbative series was
not necessary. In \cite{LLRII} it was shown that, ignoring the
non-perturbative structure function, the perturbative resummation is
crucial, and that the cut hadronic rate can not be calculated in a
systematic way. However, as was emphasized in \cite{LLRII}, this
conclusion may not stop us from extracting $|V_{ub}|$ since the
inclusion of the structure function can greatly soften the effects of
perturbative resummation.

In \cite{LLRI} it was shown how the endpoint data from radiative decay
can be used to eliminate our ignorance of the Fermi motion of the
heavy quark (i.e., the non-perturbative structure function) in the
electron end point spectrum.  In this paper we will apply these ideas
to the hadronic mass spectrum.  The next section is a synopsis of the
results given in \cite{LLRII} for the resummed hadronic invariant mass
spectrum, which is followed by a section dedicated to applying the formalism of
\cite{LLRI} to derive a closed form expression for $|V_{ub}|$ in
terms of the hadronic mass spectrum. The final section of the paper
includes the results and a discussion of the errors.

\section{Infrared Factorization and Resummation}

The factorization and resummation for inclusive $B$ decays in leptonic
variables are discussed in Ref.~\cite{KS}. It was extended to hadronic
variables in Ref.~\cite{LLRII}.  Here we will briefly review the
results.  Consider the inclusive semi-leptonic decay of the $b$ quark
into a lepton pair with momenta $q=(p_{l}+p_\nu)$ and a hadronic jet
of momenta $p_h$. We define the following leptonic kinematic variables
\begin{eqnarray}
y_0 &=& \frac{2\,v\cdot q}{m_b},\\
y &=& \frac{q^2}{m_b^2},\\
x &=& \frac{2\,v\cdot p_{l}}{m_b},
\end{eqnarray}
where $v=(1,\vec{0})$, and partonic kinematic variables
\begin{eqnarray}
\hat s_0 &=& \frac{s_0}{m_b^2} =\frac{p_h^2}{m_b^2}, \\
h &=& \frac{2\,v\cdot p_h}{m_b}.
\end{eqnarray}
In terms of the leptonic variables, $\hat s_0=(1-y_0+y)$ and
$h=2-y_0$, one can see that in the end point region of the electron
energy spectrum when $x\to 1$ with $y<1$, the invariant mass of the
jet approaches zero with its energy held fixed.  In addition, the jet
hadronizes at a much later time in the rest frame of the $B$ meson,
due to the time dilation. Factorization exploits this and separates
the particular differential rate under consideration into subprocesses
with disparate scales. This factorization fails when the jet energy
vanishes in the dangerous region $y\to x\to 1$. However, this
problematic region of phase space is suppressed because the rate to
produce soft massless fermions vanishes at tree level.

In terms of $y_0$ and $y$, the triply differential rate, which
factorizes into hard, jet and soft subprocesses \cite{KS}, may be
written as
\begin{equation}
\frac1{\Gamma_0}\frac{d^3\Gamma}{dy_0\,dy\,dx} = 
\; 12\, (y_0 - x)(x - y)
  \int_\xi^{x_m} dz\,S(z)\,m_b^2\,J[m_b^2\; h (z-\xi),\mu]\,
  H(m_b\,h/\mu),
\end{equation}
\begin{equation}
 \Gamma_0 = \frac{G_F^2}{192 \pi^3}|V_{ub}|^2 m_b^5,
\end{equation}
where $x_m=M_B/m_b$, $\mu$ is a factorization scale, and $\xi =
(1-y)/(2-y_0)$ is analogous to the Bjorken scaling variable in deep
inelastic scattering.  $z=1+k_+/m_b$, where $k_+$ is the heavy quark
light cone residual momentum. $S(z)$ essentially describes the
probability for the $b$ quark to carry light cone momentum fraction
$z$ and allows for a leakage past the partonic endpoint, as can be
seen explicitly in the upper limit of $z$.
 
It is convenient to change variables from $y$ to $\xi$ and perform the
$x$ integration to yield
\begin{equation}
\label{doubrate}
 \frac1{\Gamma_0}\frac{d^2\Gamma}{dy_0\,d\xi} = 
 2 (2 - y_0)^3 (2 y_0 - 1) \int_\xi^{x_m} dz\,S(z)\,m_b^2\,
  J[m_b^2\,(2-y_0)\,(z-\xi)]\,H[m_b\,(2-y_0)].
\end{equation}
To proceed, we take the $N$th moment with respect to $\xi$ in the
large $N$ limit. In the region $\hat s_0 \sim 0$ and $z \sim \xi \sim
1$, one can replace $J[m_b^2\,h(z-\xi)]$ in Eq.~(\ref{doubrate}) with
$J[m_b^2\,h(1-\xi/z)]$. This replacement is permissible to the order
we are working.  We then obtain
\begin{eqnarray}
\label{moment}
M_N &=& \int_0^{x_m} d\xi\;\xi^{N-1}
   \frac1{\Gamma_0}\,\frac{d^2\Gamma}{d\xi\,dy_0} \nonumber\\
&=& 2 (2-y_0)^2 (2 y_0-1)\, S_N\, 
   J_N[m_b^2\,(2-y_0)/\mu^2]\, H[m_b(2-y_0)\,/\mu] + {\cal O}(1/N),\\
 J_N(m_b^2/\mu^2) &=& m_b^2 \int_0^1 dy\,y^{N-1} J[m_b^2(1-y),\mu], \\
 S_N &=& \int_0^{x_m} dz\; z^N S(z).
\end{eqnarray}
The soft moment $S_N$ further decomposes into a perturbative piece, 
which accounts for soft gluon radiation and a
non-perturbative piece which incorporates bound state dynamics and
serves as the boundary condition for the renormalization group
equation \cite{KS}.
\begin{eqnarray}
\label{deff}
S(z)&=& \int_z^{x_m} \frac{ds}{s} f[m_b(s-1)]\, \sigma(z/s), \\
S_N &=& f_N \,\sigma_N,
\end{eqnarray}
where $f(s)=\langle B(v)|\,\bar{b}_v\, \delta(s-iD_+)\,
b_v\,|B(v)\rangle$ is the non-perturbative structure function defined
in Ref.~\cite{N}, which resums an infinite number of operators all in
the same order of $\Lambda/m_b$.  A similar expression holds for
the inclusive $B\to X_s \gamma$ in the end point region of the photon
spectrum\cite{KS}
\begin{eqnarray}
\label{srmoment}
M^\gamma_N &=& \frac1{\Gamma^\gamma_0} \int_0^{x_m} dx_\gamma\, 
  x_\gamma^{N-1} \, \frac{d\Gamma^\gamma}{dx_\gamma} \nonumber\\
 &=& S_N\,J_N^{\gamma}\, H^\gamma ,\\
\Gamma^\gamma_0 
 &=& \frac{G_F^2|V_{ts}^*V_{tb}|^2\alpha\,C_7^2\,m_b^5}{32\pi^4}.
\end{eqnarray}
Here $C_7$ is the Wilson coefficient of the $O_7$ operator evaluated
at the scale $m_b$\cite{AG} and $x_\gamma=2E_\gamma/m_b$.

In moment space the soft and jet functions \cite{KT} have been
calculated to next-to-leading (NLL) logarithmic order \cite{AR} and
are given by
\begin{eqnarray}
\sigma_N\,J_N &=& \exp [\,\log(N)\,g_1(\chi) + g_2^{\gamma}(\chi)
  +g_{sl}(\chi,y_0)] \\
\label{mr}
\sigma_N\,J_N^\gamma &=& \exp [\,\log(N)\,g_1(\chi) + g_2^\gamma(\chi)],
\end{eqnarray}
where $\chi = \alpha_s(m_b^2)\beta_0\log N$, and $g_1$, $g_2^\gamma$
and $g_{sl}$ 
can be found in \cite{LLRI}.
The hard parts are given in \cite{LLRII}:
\begin{eqnarray}
H(y_0) &=& 1-\frac{2\alpha_s}{3\pi}\left[ 4\log^2(2-y_0) 
 +\frac{8-10y_0}{2y_0-1} \log(2-y_0) +  2 {\rm Li}_2(y_0-1) + \frac52
 +\frac{2\pi^2}3\right], \\
H^\gamma &=& 1 - \frac{2\alpha_s}{3\pi}
  \left(\frac{13}{2}+\frac{2\pi^2}{3}\right).
\end{eqnarray}

To get back the physical spectra from the moment space, the inverse
Mellin transform has to be evaluated at NLL accuracy as well. If we
are willing to ignore the effects of the structure function we can get
a closed form expression for the resummed rate. To this end, we use
the results derived in the Appendix of Ref.~\cite{LLRI}
\begin{eqnarray}
\label{semispe}
\frac1{\Gamma_0}\,\frac{d^2\Gamma}{dy_0 \,d\xi}&=&
2 (2-y_0)^2 (2y_0-1)\,H(y_0)
 \nonumber \\ 
 && \times \frac{d}{d\xi} \left\{\, 
 \theta\left(1-\xi-\eta\right)\,
 \frac{ e^{l\:g_1(\alpha_s \beta_0\, l) + g_2(\alpha_s \beta_0\, l)}}
    {\Gamma\left[1-g_1(\alpha_s \beta_0\, l)-
   \alpha_s \beta_0\, l\, g_1'(\alpha_s \beta_0\, l)\right]}\right\},
\end{eqnarray}
where $l=-\log\left(-\log\xi\right) \approx -\log(1-\xi)$ and
$g_2=g_2^\gamma+g_{sl}$.  The $\theta$-function defines the
differential rates in a distribution sense, as $\eta\to 0$, and turn
the singular terms into the '$+$' distributions.

This result would only be useful in making a physical prediction if we
had a measured structure function with which we could perform a
convolution. Instead, following \cite{LLRII} we will take the inverse
Mellin transform of the ratio between the moments of the semi-leptonic
and radiative decay rates.

\section{Extraction of $|V_{\lowercase{ub}}|$}

The key observation to extract $|V_{ub}|$ is that the soft function,
Eq.~(\ref{deff}), is universal in inclusive $B$ decays. Using
(\ref{srmoment}) -- (\ref{mr}) we can rewrite (\ref{moment}) as
\begin{equation}
M_N=  2 (2-y_0)^2 (2 y_0-1)\,\frac{H(y_0)}{H^\gamma}\,
e^{g_{sl}(\chi,y_0)}\;M_N^\gamma
\end{equation}
Taking the inverse Mellin transform we arrive at
\begin{eqnarray}
\label{invm1}
\frac{\Gamma_0^\gamma}{\Gamma_0}\frac{d^2\Gamma}{dy_0\,d\xi}
 &=& 2 (2-y_0)^2 (2 y_0-1) \nonumber \\
&&\phantom{ 2 (2-y_0)^2}\times
\int_{\xi}^{1}\,\tilde{H}(y_0) 
du\,\frac{d\Gamma^\gamma}{du}\,\left(-z\frac{d}{dz}\right)
\left[\theta(1-z-\eta)\,e^{g_{sl}(\chi_z,y_0)}\right],
\end{eqnarray}
where
\begin{eqnarray}
\label{htil}
\tilde{H}(y_0) &=& \frac{H(y_0)}{H^\gamma} \nonumber \\
    &=& 1-\frac{2\alpha_s}{3\pi}\left[ 4\log^2(2-y_0) 
 +\frac{8-10y_0}{2y_0-1} \log(2-y_0) +  2 {\rm Li}_2(y_0-1)-4\right]
\\
g_{sl}(\chi,y_0)&=& \frac4{3\pi\beta_0}\log(2-y_0)\log(1-\chi), 
\end{eqnarray}
$\chi_z=-\alpha_s\,\beta_0\,\log(1-z)$ and $z=\xi/u$.  We have now
rescaled to ``physical'' variables which are normed to the meson mass
$M_B$.  We have also dropped a factor of $\bar{\Lambda}/m_b$ which would
multiply $y_0$ in these renormed variables, as it would only be
relevant in the region where factorization breaks down, $y_0\simeq2$,
which, as previously mentioned is the suppressed part of phase space.

Now to get the cut integrated rate we change the order of integration
so that integral over photon energy is the last to be performed. This
leads to a result which can most easily be integrated over the data for
the radiative decay. We find the following expression for the rate
with a cut $c = s_c/M_B^2$ on the hadronic invariant mass spectrum:
\begin{eqnarray}
\label{trip}
\delta \Gamma(c)&=& \left(\int^{\frac{1}{1+c}}_{1-\frac{\sqrt{c}}{2}}du 
\int^1_\frac{1-\sqrt{c}/2}{u} dz \int^{4uz-2}_{2-\frac{c}{1-uz}}dy_0+
\int_{\frac{1}{1+c}}^1du \int^{\frac{1}{u(1+c)}}_\frac{1-
\sqrt{c}/2}{u} dz \int^{4uz-2}_{2-\frac{c}{1-uz}}dy_0 \right.
\nonumber \\
&&+\left.
\int_{\frac{1}{1+c}}^1 du \int^1_\frac{1}
{u(1+c)} dz \int^{4uz-2}_{2-\frac{1}{uz}}dy_0\right)
\frac{\Gamma_0}{\Gamma_0^\gamma}\; I(u,y_0,z) \\
I(u,y_0,z) &=& 2 (2-y_0)^2 (2 y_0-1) 
\tilde{H}(y_0) 
\frac{d\Gamma^\gamma}{du}\,\left(-z\frac{d}{dz}\right)
\left[\theta(1-z-\eta)\,e^{g_{sl}(\chi_z,y_0)}\right].
\end{eqnarray}

We can estimate the effects of resummation on the cut rate by using a
model for the structure function. In Figure 1, we show the effects of
the resummation of the end point logs as a function of the cut using
the same model as in \cite{FN}.  We see that the effects of
resummation are small, and thus the end point logs do not form a
dominant sub-series, and the unknown piece of the two loop result
dominates the perturbative uncertainties.  Note that this is not
inconsistent with the results of \cite{LLRI}, where it was found that
the effects of resummation were large, as the structure function
mollifies the effects of the resummation.
\begin{figure}[t]
\centerline{\epsfysize=11truecm  \epsfbox{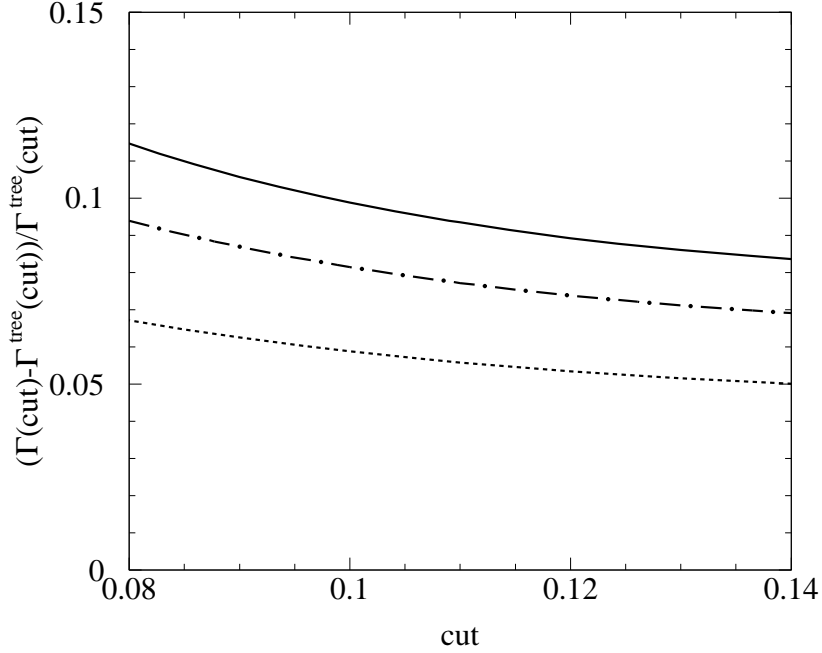} }
\tighten{
\caption{{\it Percentage change due to resummation. 
The solid line is the fully
resummed result. The dotted line is the
expansion of the resummed result to order $\alpha_s$, while the 
dot-dashed line
is the expansion of the resummed result up to order $\alpha_s^2$.}}}
\end{figure}
%


If we now expand $g_{sl}$ to order $\alpha_s$, we can write down
a simple expression for $|V_{ub}|$ as a convolution of the data 
from radiative decay. We may use the approximation
\begin{equation}
-\frac{d}{dz}\left(\theta(1-z)e^{g_{sl}}\right)\approx
\delta(1-z)+ \frac{4\alpha_s}{3\pi}\left({\log(2-y_0)\over{1-z}}\right)_+.
\end{equation}
We then find
\begin{equation}
\delta \Gamma (c)=\frac{\Gamma_0}{\Gamma_0^\gamma} [I_0(c)+I_+(c)],
\end{equation}
\begin{eqnarray}
I_0(c)&=&\left(\int^{\frac{1}{1+c}}_{1-\frac{\sqrt{c}}{2}}du 
 \int^{4u-2}_{2-\frac{c}{1-u}}dy_0+
\int_{\frac{1}{1+c}}^1 du  
\int^{4u-2}_{2-\frac{1}{u}}dy_0\right)
2(2-y_0)^2 (2y_0-1) \frac{d\Gamma^\gamma}{du}\tilde{H}(y_0),\\
\label{plus}
I_+(c)&=&\int du \int dz  \int dy_0\,  
2 (2-y_0)^2 (2y_0-1)z\, \frac{d\Gamma^\gamma}{du}\frac{4 \alpha_s}{3 \pi}
\left(\frac{\log(2-y_0)}{1-z}\right)_+.
\end{eqnarray}
The plus distribution is defined as
\begin{equation}
\int_a^1 dx\frac{f(x)}{(1-x)_+}=\int_a^1dx\frac{f(x)-f(1)}{1-x} - 
\int_0^adx\frac{f(1)}{1-x}.
\end{equation}

The regions of integration in Eq.~(\ref{plus}) are the same as those in
Eq.~(\ref{trip}).  The integrals over $y$ and $z$ can all be done
analytically. The final result is rather large and complicated, so we
choose not to include it in print. The resulting expression is an
integral which involves a known function of $u$ and $c$ and the
radiative decay endpoint data. Thus we may write
\begin{equation}
\frac{|V_{ub}|}{|V_{ts}|}=\left\{\frac{6\,\alpha\,C_7(m_b)^2\,\delta\Gamma(c)}
{\pi\,[I_0(c)+I_+(c)]}\right\}^{\frac12}.
\end{equation} 
The dominant source of error, modulo the usual uncertainties inherent in
{\it all} inclusive predictions, will come from unknown effects
at order $\alpha_s^2$, and $\Lambda^2/(c\, m_b^2)$, each of which should 
contribute only at the few percent level.  

\section{Results, Conclusions and Outlook}
The extraction of $|V_{ub}|$ using the cut hadronic mass spectrum has
the theoretical advantage over the lepton spectrum since the mass cut
includes a larger fraction of the $b\rightarrow u$ transition.  In
addition, we expect local parton-hadron duality to work better since
more resonances will contribute to the cut mass spectrum. The real
issue for the viability of this method is the resolution.  In
particular, how good will the ``best'' resolution be for the invariant
mass? This will be limited by the inability to detect neutrals, as
well as the charge particle detection inefficiencies. Presently, the
resolution is not good enough to eliminate the charmed background,
which has to be modeled. Hopefully, it will be possible to eventually
push the cut further\cite{T}.  If the resolution does become good
enough to really eliminate the charmed background, then there is an
additional option available.  It has been pointed out that the
leptonic mass spectrum is effectively less sensitive to the Fermi
motion of the $b$ quark \cite{Buc}. Recently, this fact has been used
to determine an expression for $|V_{ub}|$ in terms of the cut lepton
mass spectrum, without needing to use the data from radiative
decay\cite{BLL}.  However, the resulting expansion parameter is
$\Lambda/M_D$, and grows as the cut is raised above $(M_B-M_D)^2$. The
authors in \cite{BLL} calculated the leading non-perturbative
correction and found it to be anomalously small. They made an
estimate for the sub-leading corrections and found that they grow
rapidly as the cut is raised from its smallest possible value.

In the end it is clear that no one single extraction of $|V_{ub}|$
should be trusted, given the fact that our expansion parameters are
never as small as we would like. We will only gain confidence in the
extractions after there is convergence among several independent
extractions.

\acknowledgments 
This work was supported in part by the Department of Energy under
grant numbers DOE-ER-40682-143 and DE-FG03-92-ER 40701,
 and also by the NSF under grant  number 
PHY94-07194. The authors  would like to thank 
the theory group at Caltech for their hospitality. 
I.\,L. is supported
in part by an ITP Graduate Fellowship.

\tighten


\end{document}